\documentstyle[12pt]{article}
\input psfig

\def\GeV{\,{\rm GeV}}

\def\MeV{\,{\rm MeV}}

\def\rcm{\,{\rm cm}}

\def\Mpc{\,{\rm Mpc}}

\def\eV{{\,\rm eV}}

\def\cmm2{{\,\rm cm^{-2}}}
\def\cm2{{\,{\rm cm}^2}}
\def\cmm3{{\,{\rm cm}^{-3}}}
\def\gcmm3{{\,{\rm g\,cm^{-3}}}}
\def\kms{\,{\rm km\,s^{-1}}}

\def\la{\mathrel{\mathpalette\fun <}}
\def\ga{\mathrel{\mathpalette\fun >}}
\def\fun#1#2{\lower3.6pt\vbox{\baselineskip0pt\lineskip.9pt
  \ialign{$\mathsurround=0pt#1\hfil##\hfil$\crcr#2\crcr\sim\crcr}}}

\begin{document}
\pagestyle{empty}
\begin{center}
\bigskip

\rightline{FERMILAB--Pub--97/186-A}
\rightline{astro-ph/9706069}
\rightline{{\it submitted to Reviews of Modern Physics (Colloquia)}}

\vspace{.2in}
{\Large \bf Big-Bang Nucleosynthesis \\
\bigskip
Enters the Precision Era}

\bigskip

\vspace{.2in}
David N. Schramm and Michael S. Turner\\

\vspace{.2in}
{\it Departments of Physics and of Astronomy \& Astrophysics\\
Enrico Fermi Institute, The University of Chicago, Chicago, IL~~60637-1433}\\

\vspace{0.1in}
{\it NASA/Fermilab Astrophysics Center\\
Fermi National Accelerator Laboratory, Batavia, IL~~60510-0500}\\

\end{center}

\vspace{.3in}
\centerline{\bf ABSTRACT}
\bigskip
The last parameter of big-bang nucleosynthesis,
the baryon density, is being pinned down by measurements
of the deuterium abundance in high-redshift hydrogen
clouds.  When it is determined, it will fix the primeval light-element
abundances.  D, $^3$He and $^7$Li will become ``tracers'' for the study of
Galactic and stellar chemical evolution, and big-bang nucleosynthesis
will become an even sharper probe of particle physics, e.g., the
bound to the number of light neutrino species will be tightened significantly.
Two key tests of the consistency of the standard theory are
on the horizon:  an independent, high-precision determination of
the baryon density from anisotropy of the cosmic background radiation and
a precision determination of the primeval $^4$He abundance.

\newpage
\pagestyle{plain}
\setcounter{page}{1}
\newpage

\section{From Gamow to Keck}

Over the last two decades big-bang nucleosynthesis (BBN) has emerged as one of
the cornerstones of the big bang, joining the Hubble expansion and the
Cosmic microwave Background Radiation (CBR) in this role.
Of the three, big-bang nucleosynthesis probes the Universe to the
earliest times, from a fraction of a second to hundreds of seconds.
Since BBN involves events that occurred at temperatures of order 1\,MeV,
it naturally played a key role in forging
the connection between cosmology and nuclear and particle physics
that has blossomed during the past fifteen years.

It is the basic consistency of the predictions for the abundances of
the four light-elements D, $^3$He, $^4$He and $^7$Li with their measured
abundances, which span more than nine orders of magnitude,
that moved BBN to the cosmological centerstage.
In its success, BBN has led to the most accurate determination of the
mass density of ordinary matter:  Consistency holds only if
the fraction of critical density contributed by baryons ($\equiv \Omega_B$)
is between $0.007h^{-2}$ to $0.024h^{-2}$.  This ``measurement'' has
three important implications for cosmology:  baryons cannot close the
Universe; most of the baryons are dark; and most of the matter
is nonbaryonic.  (The Hubble constant $H_0=100h\kms\Mpc^{-1}$
enters because it fixes the critical density; recent measurements
seem to be converging on a value $h=0.65\pm 0.1$.)

Currently, there is great excitement because we are on the verge of
determining the baryon density to a precision of 20\%
or better from measurements of the primeval deuterium abundance.
When this occurs, BBN will enter a qualitatively new phase --
an era of high precision.  The consequences for cosmology are
clear -- pinning down the baryon density and completing the
story of BBN.  The implications for astrophysics are just as important --
fixing the baryon density fixes the primeval abundances of the light elements
and allows them to be used as tracers in the study of
the chemical evolution of the Galaxy and aspects of stellar evolution.
Finally, important limits to particle properties, such
as the limit to the number of light neutrino species,
can be further sharpened.

The BBN story \cite{bbnstory} begins with Gamow and his collaborators,
Alpher and Herman, who viewed the early Universe as a nuclear furnace that
could ``cook the periodic table.''  Their speculations, while
not correct in all details, led to the prediction of the CBR.
Key refinements include those made by Hayashi recognized the
role of neutron-proton equilibration, and by Turkevich and Fermi
pointed out that lack of stable nuclei of mass 5 and 8 precludes
light nucleosynthesis beyond the lightest elements.
The framework for the calculations themselves
dates back to the work of Alpher, Follin and Herman and of Taylor
and Hoyle, preceding the discovery of the 3K background,
of Peebles and of Wagoner, Fowler and Hoyle,
immediately following the discovery, and the more recent work of our
group of collaborators \cite{chicagogroup} and of other groups
around the world \cite{others}.

\begin{figure}[t]
\centerline{\psfig{figure=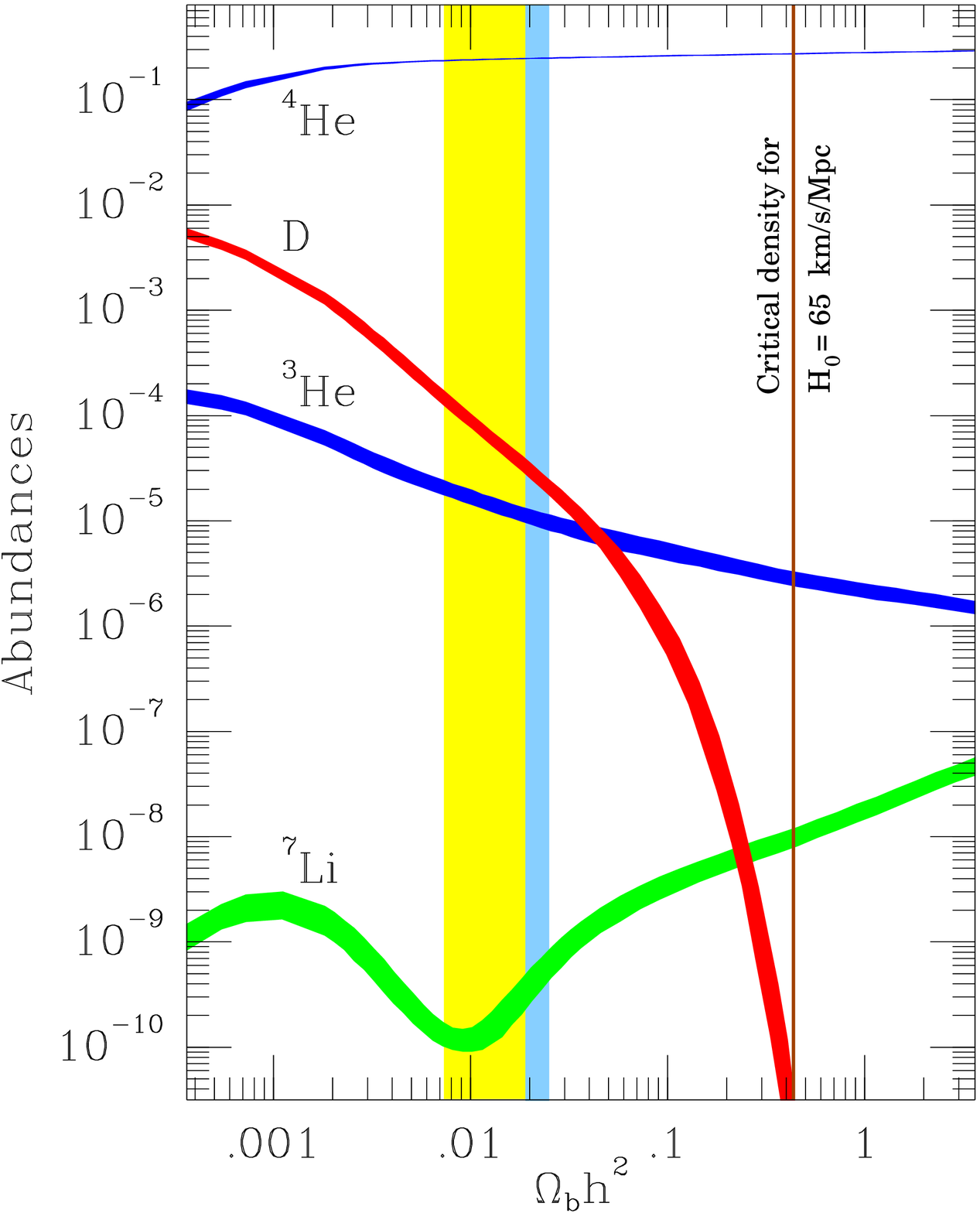,width=5in}}
\caption{Summary of big-bang production of the light elements.  The widths
of the curves indicate the $2\sigma$ theoretical uncertainties,
and the vertical band is the Copi et al \protect\cite{cst}
consistency interval where the predicted abundances of
all four light elements agree with their measured primeval
abundances.  The darker band in the consistency interval corresponds
to Tytler et al's determination of the primeval deuterium abundance
(Figure courtesy of K.~Nollett).}
\label{fig:newbbn}
\end{figure}

The basic calculation, a nuclear reaction network in an expanding
box, has changed very little.  The most up to date
predictions are shown in Fig.~\ref{fig:newbbn}.
The predictions of BBN are robust because essentially all
input microphysics is well determined:
The relevant energies, 0.1 to 1\,MeV, are explored in
nuclear-physics laboratories and the experimental uncertainties
are minimal, though not unimportant (see Fig.~\ref{fig:newbbn}).

Over the last twenty-five years the focus has been on understanding
the evolution of the light-element abundances from the big bang
to the present in order to test the BBN predictions for the {\it primeval}
abundances.  (Astronomers refer to the evolution of the elemental abundances
due to nuclear transmutations as ``chemical evolution.'')
In the 1960s, the main focus was $^{4}$He,
which is very insensitive to the baryon density.  The agreement between the
BBN prediction -- lots of $^4$He production -- and observations/chemical
evolution -- observed $^4$He abundance, 25\% to 30\%, is
much greater than what stars can make, a few percent -- gave strong
support to the big-bang model but gave no
significant constraint to the baryon density.

During the 1960s, there was little cosmological interest in the other
light isotopes, which are, in principle, capable of
giving information about the baryon density, because they were
assumed to have been made during the T-Tauri phase of
stellar evolution \cite{fow62}.  That changed in the 1970s and
primordial nucleosynthesis developed into
an important probe of the Universe.  In part, this was
stimulated by Ryter et al \cite{ryt70} who showed that the T-Tauri
mechanism for light-element synthesis failed.  Furthermore, knowledge
of the deuterium abundance improved significantly with solar-wind
and meteoritic measurements \cite{gei71,bla71} and
the interstellar medium (ISM) measurements
made by the Copernicus satellite \cite{rog73}.

Reeves, Audouze, Fowler and Schramm \cite{ree73} argued for a
cosmological origin for deuterium.  By exploiting the rapid
decline in deuterium production with baryon density (D/H $\propto
1/\rho_B^{1.7}$) they were able to place an upper
limit to the baryon density which excluded a Universe closed
by baryons.  This was the beginning of the use of deuterium as
a cosmic baryometer, which will soon culminate in an accurate
determination of the baryon density.  Their argument was strengthened when
Epstein, Lattimer and Schramm \cite{eps76} showed conclusively
that no realistic astrophysical process other than the big bang
could produce significant deuterium (most astrophysical processes
destroy deuterium because it is so weakly bound), and thus, the contemporary
abundance leads to a firm upper limit to the baryon density.

In the late 1970s, attention turned to $^{3}$He.  In part, this
was to exploit the steep dependence of deuterium production upon
the baryon density to constrain it from below by using the fact
that deuterium is burned to $^3$He.  In particular, it was argued
that $^3$He, unlike D, is made in stars:  during the pre-main-sequence
stage by burning deuterium and in low-mass stars during the main
sequence stage.  Thus the sum D+$^3$He should increase with time
or at least stay constant \cite{ytsso}.
Unfortunately, this simple argument is not correct in detail.
A recent measurement of $^3$He in the local ISM \cite{glo96}
has shown that D+$^3$He has been constant for the last 5 Gyr,
contradicting a significant increase
due to $^3$He production by low-mass stars, and further, the
$^3$He abundance within the Galaxy shows great variation.
The chemical evolution of $^3$He is not fully understood; however,
because the only stars that efficiently destroy $^3$He are massive
and also make metals, metal production provides an upper
limit to the amount by which D+$^3$He can decrease and
thus a lower bound to the baryon density \cite{cst2}.

The abundances of D, $^3$He and to a lesser extent $^4$He
led to the prediction that the primeval $^{7}$Li abundance should
be near its minimum, $^{7}{\rm (Li/H)} \sim 10^{-10}$.
This was verified by Spite and Spite
\cite{spite2}, who measured the $^7$Li abundance in the
atmospheres of the oldest (pop II) stars in the halo of
our galaxy.  Their work was confirmed and extended by Hobbs,
Thorburn, and others \cite{thorburnetal}.  An important
question still remains --
could the $^7$Li abundance in these stars have been reduced
by nuclear burning over the past 10 Gyr or so?

The status of BBN was reviewed and summarized in 1995
by Copi et al \cite{cst} who concluded:
Within the uncertainties -- chemical evolution for $^3$He and D, stellar
depletion for $^7$Li and systematic error
for $^4$He -- the abundances of the four light elements
produced in the big bang are consistent with their BBN predictions
provided that the fraction of critical density contributed
by baryons is between $0.007h^{-2}$ and $0.024h^{-2}$ and
the equivalent number of light neutrino species is less than 3.7.
This is an impressive achievement; it will be eclipsed
when the full potential of deuterium as the cosmic baryometer is realized.

\section{Keck:  The Great Leap Forward}

As discussed above, it took a while to recognize the cosmic importance of
deuterium and its role as the baryometer.  Measuring the primeval
deuterium abundance has take even longer and
required the advent of the 10 meter W.M. Keck Telescope
and its HiRes spectrograph.  However, it was worth the wait.

In 1976 Adams \cite{adams} outlined how the deuterium abundance
in a high-redshift hydrogen cloud could be measured.
Distant hydrogen clouds are observed in absorption against
even more distant quasars.  Many absorption features are seen --
the Lyman series of hydrogen and the lines of various ionization
states of carbon, oxygen, silicon, magnesium, and other elements.
Because of the large hydrogen abundance, Ly$\alpha$
is very prominent.  In the rest
frame Ly$\alpha$ occurs at $1216$\AA, so that for a
cloud at redshift $z$ Ly$\alpha$ is seen at
$1216(1+z_{\rm cloud})$\AA.  The isotopic shift for deuterium
is $-0.33(1+z)$\AA, or expressed as a Doppler velocity, $-82\kms$.
Adams' idea was to detect the deuterium Ly$\alpha$
feature in the wing of the hydrogen feature.  (The same
technique is used to detect deuterium in the local ISM.)

His proposal has much to recommend it:  For $z\ga 3$,
Ly$\alpha$ is shifted into the visible part of the spectrum
and thus can be observed from Earth;
``Ly$\alpha$ clouds'' are ubiquitous with hundreds
being seen along the line of sight to a quasar of this redshift, and
judged by their metal abundance, anywhere from $10^{-2}$
of that seen in solar system material to undetectably small levels,
these clouds represent nearly pristine samples of cosmic
material.  There are technical challenges:
Because the expected deuterium abundance is small,
D/H$\,\sim 10^{-5}-10^{-4}$,
clouds of very high column density, $n_H \ga 10^{17}\rcm^{-2}$, are
needed; because hydrogen
clouds are ubiquitous, the probability of another, low column-density
cloud sitting in just the right place to mimic deuterium -- an
interloper -- is not negligible; many clouds have broad absorption
features due to large internal velocities or complex velocity structure;
and to ensure sufficient signal-to-noise bright QSOs and large-aperture
telescopes are a must.  Based upon his experience, Tytler has estimated
that no more than one in thirty quasars has a cloud suitable for
determining the primeval deuterium abundance.

Since the commissioning of the HiRes spectrograph on Keck-I,
a number of detections, tentative detections,
upper limits and lower limits for the primeval deuterium --
not all consistent with one another --
have been reported \cite{dlist}.  However, a confusing situation
is now becoming clear.  Tytler and his collaborators \cite{tytleretal}
have made a strong case for a primeval deuterium abundance of
(D/H)\,$=(2.7\pm 0.3)
\times 10^{-5}$, based upon two clouds.  One of the clouds is at
redshift 3.572 along the line of sight to quasar Q1937-1009; the other is at
redshift 2.504 along the line of sight to quasar Q1009+2956.  The metal
abundances in these clouds are around $10^{-3}$ of solar, so that any depletion
of deuterium due to stellar processing should be negligible.  In addition,
they have observed the clouds for which others had claimed a much
higher abundance, and, with better data, they have shown that
the absorption features are not deuterium \cite{nullify}.

\begin{figure}[t]
\centerline{\psfig{figure=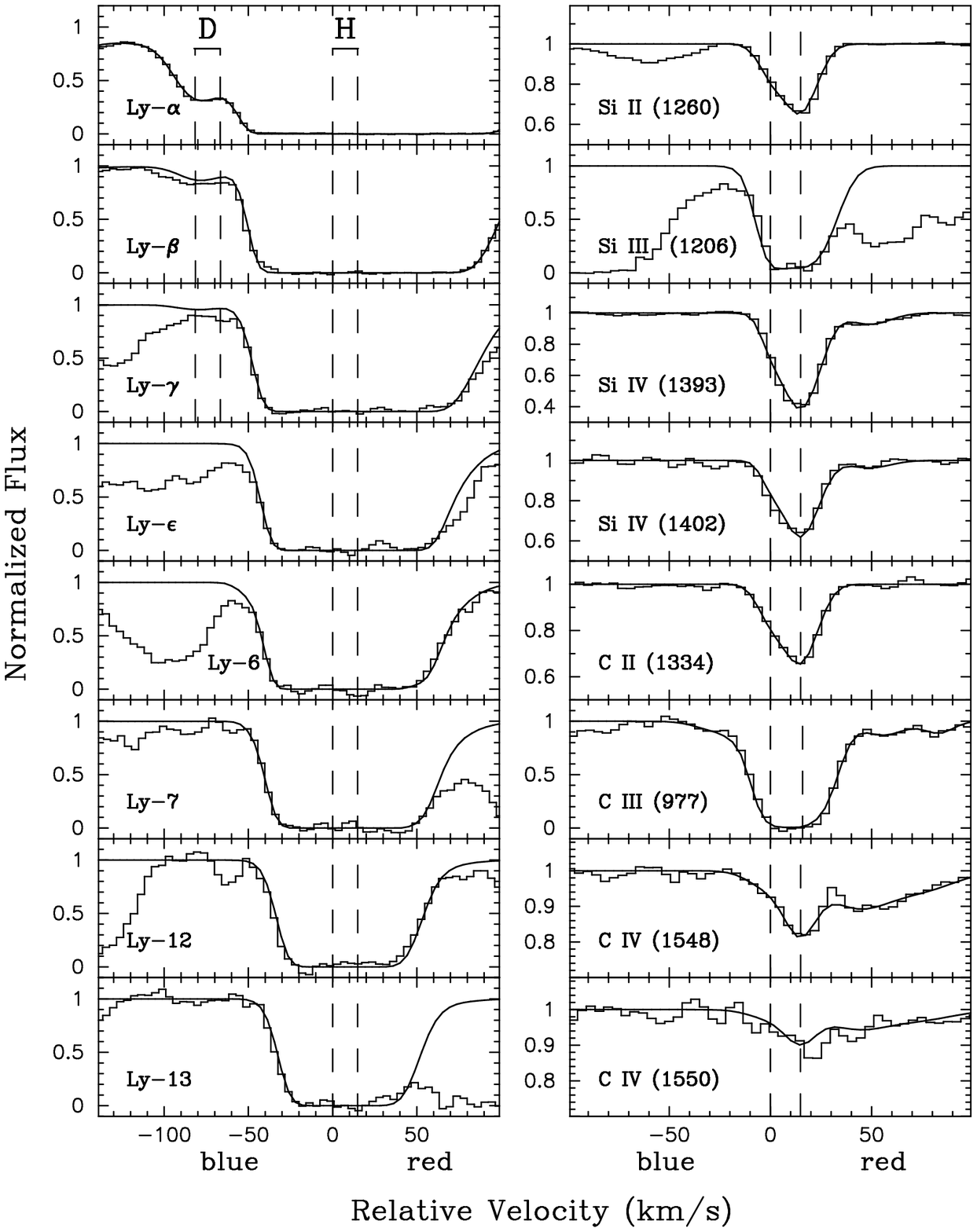,width=5in}}
\caption{Portion of the Keck HiRes spectrum of Q1937-1009.
Solid line indicates the model fit and the deuterium Ly-$\alpha$ feature is
indicated.  The left panels show the Lyman series for the cloud
and the right panels show the narrow metal lines associated with
the cloud, which are crucial to determining the positions
of the two components of the cloud (from Ref.~\protect\cite{tytleretal}).}
\label{fig:keck}
\end{figure}

It would be premature to conclude that the value of the
primeval deuterium abundance has been completely settled,
or that all potential systematic errors are fully understood \cite{cowie}.
For example, because the hydrogen Ly$\alpha$ feature is so saturated,
it is the hydrogen abundance, not deuterium, that is most difficult
to determine;
in addition, the clouds usually consist of more than one velocity
component, which complicates the analysis.  However, Tytler and his
collaborators have made a strong case for a primeval deuterium
abundance of (D/H)\,$\simeq (2.7\pm 0.3) \times 10^{-5}$, with a
possible systematic uncertainty of comparable size.
The case will be made very firm when a few
more clouds of similar deuterium abundance are found, or conversely,
the case could fall apart.
Both Keck and HST observations are ongoing.  The UV capability of HST
allows a search at lower redshift where there are fewer clouds
and the problem of interlopers mimicking deuterium is less severe.

\section{The Baryon Density and Its Cosmic Implications}

To be definite and to allow for possible systematic uncertainty,
we take as a provisional primeval deuterium
abundance, (D/H)$_P = (2.7\pm 0.6) \times 10^{-5}$.
This pegs the baryon density at $(4.0\pm 0.8)\times 10^{-31}\gcmm3$,
or as a fraction of critical density, $\Omega_B = (0.022\pm 0.004)h^{-2}$
(theoretical uncertainty included).  This lies near the high end
of the pre-Keck BBN concordance interval
and narrows the range considerably.

This big-bang determination of the baryon density
is consistent with other, independent
methods:  (1) The density of baryons in
gas at redshifts between two and four is constrained by the measured
Ly$\alpha$ opacity of the ubiquitous hydrogen clouds previously discussed and
the baryon density inferred by this method is
$\Omega_{\rm gas} \simeq (0.01 - 0.02)h^{-2}(h/0.65)^{1/2}$ \cite{lyaomegab}.
(2) Most of the baryons in clusters
of galaxies exist in the form of hot x-ray emitting gas.
Assuming that galaxy clusters represent
a fair sample of material in the Universe, the
cluster baryon fraction, which is determined from x-ray measurements
to be $f_B = (0.07 \pm 0.007)h^{-3/2}$ \cite{evrard}, can be used
to infer the universal baryon density $\Omega_B$ from
the matter density $\Omega_M$:
\begin{equation}
{\Omega_B\over \Omega_M} = f_B \ \ \ \ \Rightarrow \ \ \ \
\Omega_Bh^2  =  (0.017 \pm 0.002)(h/0.65)^{1/2}(\Omega_M / 0.3)\,.
\end{equation}
(3) The height of the Doppler peak in the angular power spectrum
of CBR anisotropy depends the baryon density (see Fig.~\ref{fig:baryoncbr});
while the data do not yet determine the baryon density very
precisely, they are consistent with the BBN value.

\begin{figure}[t]
\centerline{\psfig{figure=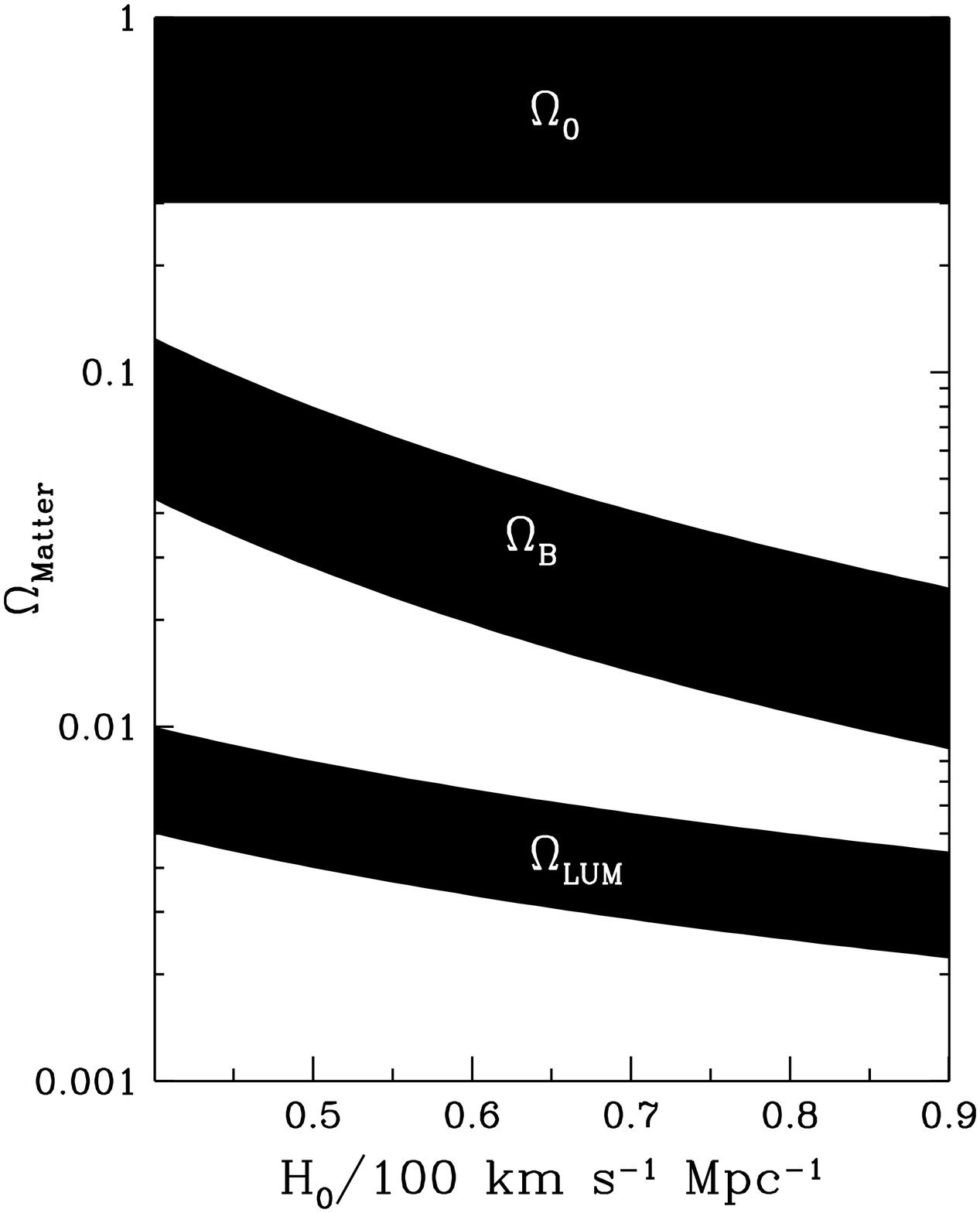,width=5in}}
\caption{Summary of knowledge of the matter density $\Omega_M$.
The lowest band is luminous matter, in the form of bright
stars and associated material; the middle band is the pre-Keck
big-bang nucleosynthesis concordance interval;
the upper region is the estimate of the total matter density
based upon dynamical methods \protect\cite{omegamatter}.
The gaps between the bands illustrate
the two dark matter problems: most of the ordinary matter is dark
and most of the matter is nonbaryonic.}
\label{fig:omegamatter}
\end{figure}

Next, consider the implications of the nucleosynthesis determination
of the baryon density.  First and foremost, it is the linchpin
in the case for the two dark matter problems central to astrophysics
and cosmology.
\begin{enumerate}

\item The big-bang determination
together with measurements of the total amount of matter, provide
firm evidence for nonbaryonic dark matter (see Fig.~\ref{fig:omegamatter}).
Dynamical measurements of the density of matter that clusters,
based upon galaxy-cluster mass determinations,
measurements of peculiar velocities, and the frequency of gravitational
lensing, indicate that $\Omega_M$ is at least 0.3
\cite{omegamatter}; nucleosynthesis puts the
baryonic contribution at a value far below, $(0.052\pm 0.01)(0.65/h)^2$.
Particle physics provides three compelling candidates for the
nonbaryonic matter:  a very light axion (mass $\sim 10^{-5}\eV$);
a light neutrino species (mass $\sim {\cal O}(10\eV )$); and the
lightest supersymmetric particle (neutralino of mass $30\GeV$ to
$500\GeV$).  That most of the matter is nonbaryonic receives
additional support:  There is no model for the formation of structure
without nonbaryonic matter that is consistent with
the anisotropy of the CBR.

\item The BBN determination also implies that most of the baryons
are in a form yet to be identified.  Stars and closely related material
(``luminous matter'') contribute less than 1\% of the critical
density, $\Omega_{\rm LUM} \simeq 0.003h^{-1}$;  since this is almost
a factor of ten lower than the BBN determination of the
baryon density, it follows that
most of the baryons are not optically bright (``dark'').

\end{enumerate}

The fact that the fraction
of critical density in gas at redshifts two to four and in
gas at the time of formation of clusters, redshifts one or less,
is consistent with the nucleosynthesis value for the baryon density,
suggests that the bulk of the  ``dark'' baryons are diffuse, hot gas.
In clusters, this is clear -- most of the baryons
are in the hot intracluster gas that shines bright in x ray.
Individual galaxies have shallower potential wells and
the gas would have a temperature of only around $10^5\,$K,
making it difficult to detect.  There is some
evidence, absorption of quasar light by singly ionized helium, for
diffuse, intergalactic gas \cite{davidson}.
While most of the dark baryons are likely in the
form of diffuse, hot gas, some fraction of the dark baryons, perhaps
10\%, are likely to be in the form of dark stars (or MACHOs),
e.g., white dwarfs, neutron stars, brown dwarfs and so on.
There is evidence from microlensing that dark stars comprise
a portion of the the halo of our own galaxy \cite{MACHOS}.

Turning a previous argument around, accepting the baryon density based
upon the primeval deuterium abundance, the cluster baryon fraction
can be used to infer the matter density:
\begin{equation}
\Omega_M = \Omega_B /f_B = (0.35\pm 0.1)(0.65/h)^{1/2}.
\end{equation}
Taken at face value, this implies that the matter density,
while much larger than the baryon density, is far from unity.
(This technique is not sensitive to a smooth component such as
vacuum energy, and does not preclude $\Omega_0 =1$ with
$\Omega_{\rm VAC} \sim 0.65$, or an even more exotic smooth component.)
However, important assumptions underlie the determination of
the cluster baryon fraction:
the gas is supported by its thermal motions only
and not magnetic fields or bulk motion; the gas is not clumped;
and clusters provide a fair sample of matter in the Universe.
If any one of these assumptions is not valid, the cluster gas
fraction would be lower and the estimate for $\Omega_M$
correspondingly higher.  There is some evidence that this
may be the case -- cluster masses
determined by gravitational lensing appear to be systematically
larger than those determined by x-ray measurements \cite{clustermass},
perhaps as much as a factor of two.

\section{Nuclear Cosmology Clarifies Galactic Chemistry}

Chemical evolution issues have been
interwoven into the study of BBN from the start.  In order to extrapolate
contemporary abundances to primordial abundances the
use of stellar and Galactic chemical-evolution models is unavoidable.
The difficulties are well illustrated by $^3$He:  generally the idea
that the sum D+$^3$He is constant or slowly increasing seems to be true,
but the details, e.g., predicted increase during the last few Gyr,
are inconsistent with a measurement of the $^3$He abundance in the local ISM.

The pinning down of the baryon density turns the tables around.
Primeval abundances become fixed and comparison with
contemporary abundances can be used to reveal the details of stellar and
Galactic chemical evolution.  Nuclear physics in the early Universe
provides tracers to study Galactic chemistry!  For sake of illustration
we continue to our provisional baryon density, $(4.0\pm 0.8)\times
10^{-31}\gcmm3$, and remind the reader that conclusions could change if
the value for the primeval deuterium abundance changes.

Beginning with deuterium, our assumed primeval abundance,
D/H $= (2.7\pm 0.6) \times
10^{-5}$, is not quite a factor of two larger than the present ISM
abundance, D/H $=(1.5 \pm 0.1) \times 10^{-5}$, determined by
Hubble Space Telescope observations \cite{linsky}.  This implies 1)
little nuclear processing over the history of the Galaxy; and/or
2) significant infall of primordial material into the disk
of the Galaxy.  The metal composition of the Galaxy, which indicates
significant processing through stars, together
with the suggestion that even more metals may have been made
and ejected into the IGM (apparently this happens at least in clusters
of galaxies), means that option one is less
likely than option two.  Even more intriguing is the fact that
the inferred abundance of deuterium in the pre-solar nebula, D/H $=
(2.6\pm 0.4)\times 10^{-5}$ \cite{presolar},
indicates less processing in the first
10 Gyr of Galactic history than in the past 5 Gyr, perhaps
suggesting a decreasing rate of infall and/or a change in the
distribution of stellar masses.

Moving on to $^3$He, the primeval value corresponding to our
assumed deuterium abundance is $^3$He/H $\simeq 10^{-5}$.
The pre-solar value, measured in meteorites and more recently in the outer
layer of Jupiter, is $^3$He/H $=(1.2 \pm 0.2)\times 10^{-5}$ \cite{presolar},
comparable to the primeval value.  The value in the present ISM,
$^3$He/H $=(2.1\pm 0.9)\times 10^{-5}$, is
about twice as large as the primeval value \cite{glo96}.
On the other hand, the primeval sum of deuterium and
$^3$He, $(3.7\pm 0.7)\times 10^{-5}$ relative to H, is essentially equal
to that determined for the pre-solar nebula, $(3.8\pm 0.4)\times
10^{-5}$, and for the present ISM, $(3.7\pm 1)\times 10^{-5}$.
This indicates little net $^3$He production beyond the burning
of deuterium to $^3$He, and conflicts with conventional models for
the evolution of $^3$He which predict a significant increase in D+$^3$He
(due to $^3$He production by low-mass stars), as well as unconventional
models where $^3$He is efficiently burned.  The constancy of D+$^3$He
might be actually be a coincidence:  In models put forth
to explain certain isotopic anomalies ($^{18}$O/$^{16}$O
and $^{12}$C/$^{13}$C) \cite{wasserburg}, $^3$He is
produced during the main-sequence phase and then destroyed during
post-main-sequence evolution.  While empirical evidence supports
the idea that the D+$^3$He remains roughly constant, clearly
much remains to be learned about the chemical evolution of $^3$He.

Finally, consider $^7$Li.  The predicted primeval abundance, $^7$Li/H $=(5\pm 2)
\times 10^{-10}$, is a factor of two to three larger than that
measured in the atmospheres of pop II halo stars, $^7$Li/H $=(1.5\pm 0.3)
\times 10^{-10}$.  There are two plausible explanations.  The abundance
determinations in these old halo stars are sensitive to the
model atmospheres used, and there could be as much as 50\%
uncertainty due to this.  Or lithium could have been depleted in
these old stars.  The observation of
$^6$Li in at least one pop II halo, which is much more fragile
than $^7$Li, limits stellar depletion to a factor of two or less
\cite{lemoineetal}; further, the fact that the $^7$Li abundance
is at most weakly dependent upon stellar mass also argues for
a depletion of at most a factor of two or so \cite{li7deplete}.
(Both arguments seem to be validated since the ratio of the pop II abundance
to primeval abundance is about a factor of two.)
If depletion is important, as seems likely,
further, high-quality observations
of these old halo stars should begin to reveal dispersion
in the $^7$Li abundance due to the difference in rotation rates
and/or ages.  Tidally locked binaries, where the rotation rate is known,
are especially useful.  When the final details of the $^7$Li story
are in, much will have been learned about the role of rotation
and mixing in stellar evolution.

\section{Helium-4:  A Loose End}

\begin{figure}[t]
\centerline{\psfig{figure=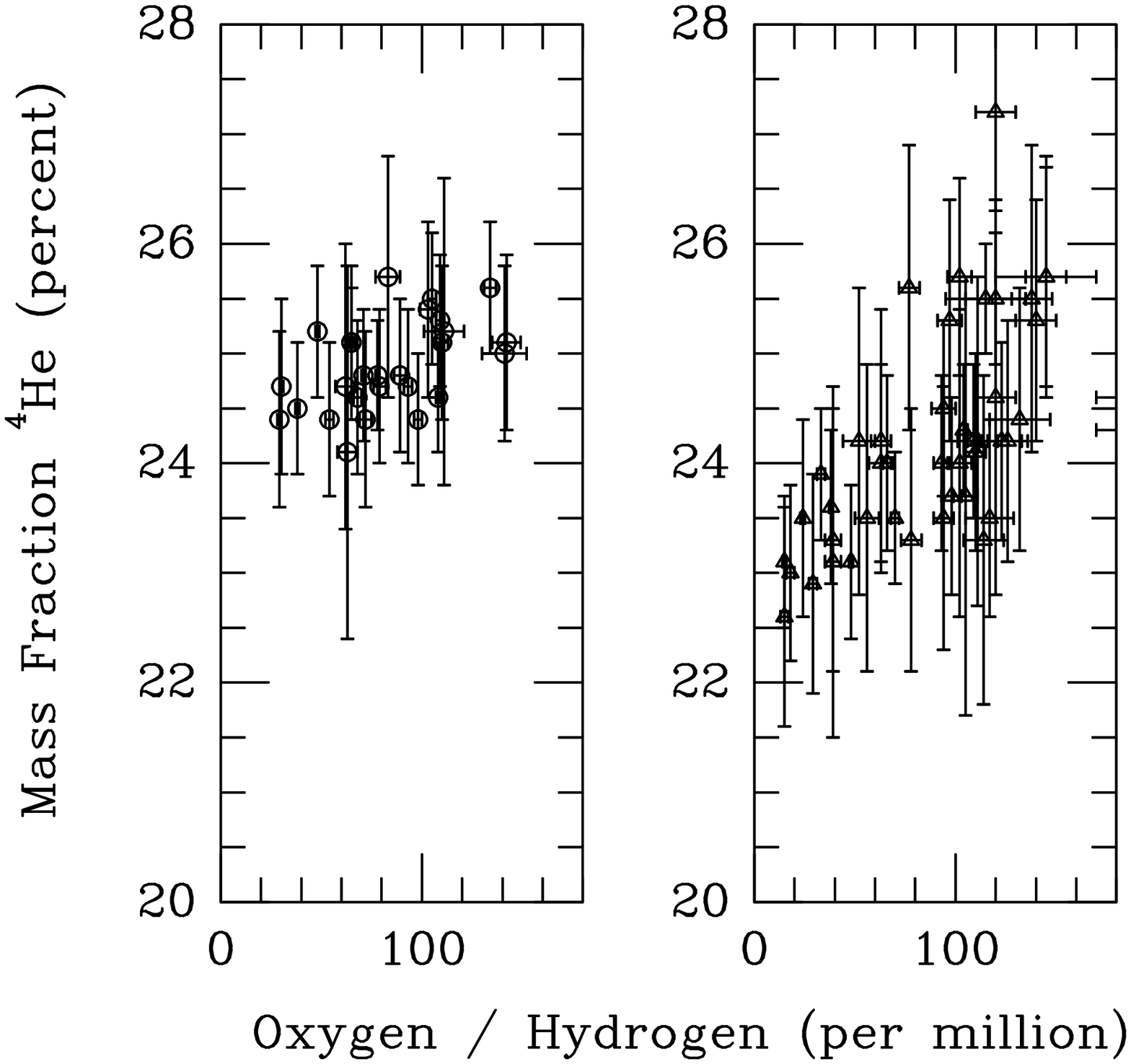,width=5in}}
\caption{Helium-4 abundance vs. oxygen abundance
in metal-poor, dwarf emission-line
galaxies.  Right panel (triangles) is the sample analyzed by
Olive and Steigman \protect\cite{os}; left panel (circles)
is the new sample of Izotov et al \protect\cite{izotovetal}.}
\label{fig:he4}
\end{figure}

\begin{figure}[t]
\centerline{\psfig{figure=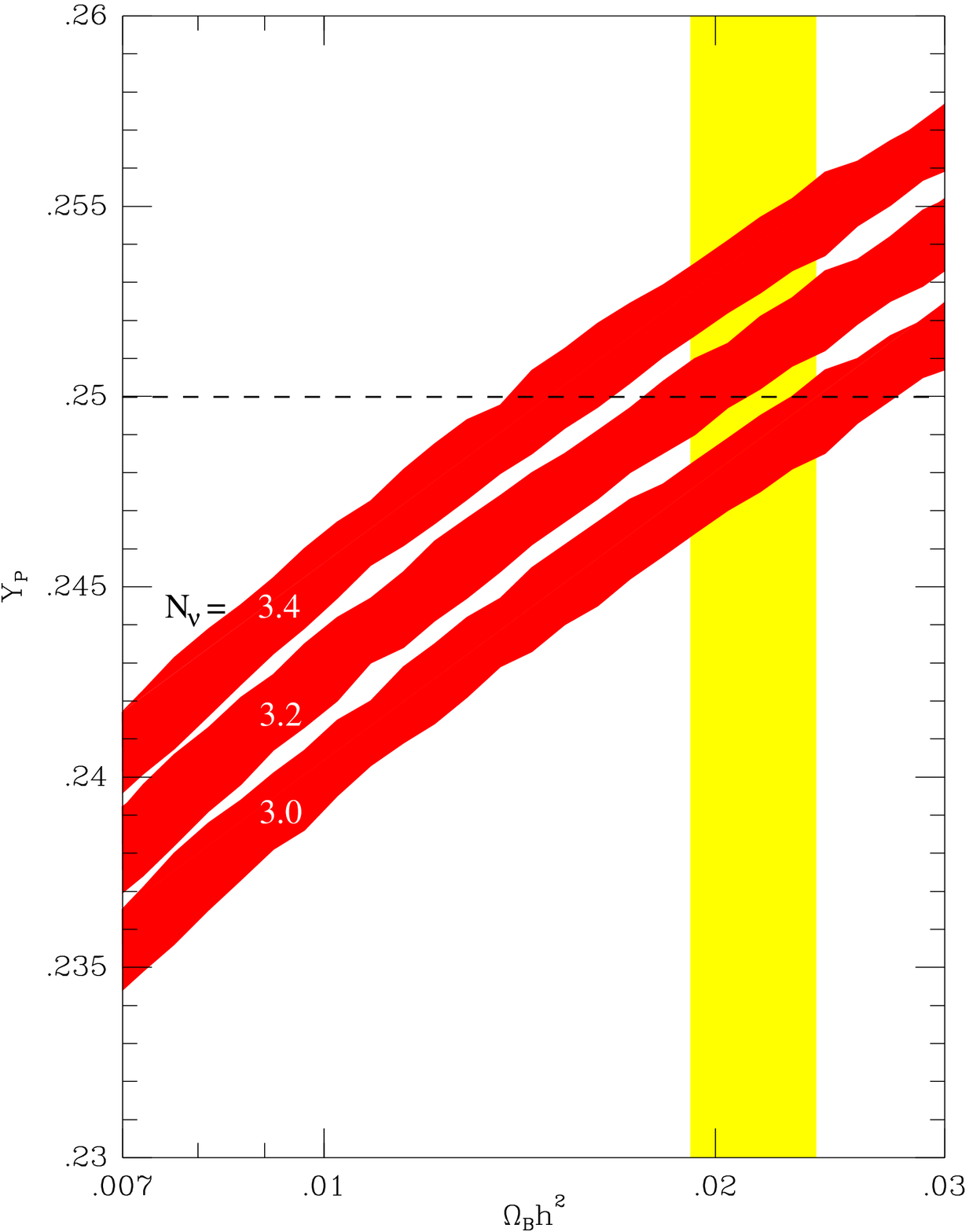,width=5in}}
\caption{$^4$He production for $N_\nu =3.0, 3.2, 3.4$.
The vertical band indicates the baryon density
consistent with (D/H)$_P = (2.7\pm 0.6) \times 10^{-5}$ and the horizontal
line indicates a primeval $^4$He abundance of 25\%.
The widths of the curves indicate the two-sigma theoretical uncertainty.}
\label{fig:bbnbig}
\end{figure}

Helium-4 plays a different role and presents different challenges.
First, the primeval yield of $^4$He is relatively insensitive to the
baryon density -- pinning down the baryon density to 20\% pegs
its value to 1\% precision (see Fig.~\ref{fig:bbnbig}). Secondly,
the chemical evolution of $^4$He is straightforward --
the abundance of $^4$He slowly increases due to stellar production.
Helium-4 was the first important test of BBN;
a high-precision determination of the primeval $^4$He abundance
will guarantee that it has an important future role too.

Here is the present situation:  Assuming our provisional value
for the primeval deuterium abundance, the predicted primeval
$^4$He abundance is $Y_P=0.2475 \pm 0.002$ (including theoretical uncertainty).
There have been two recent determinations of the
primeval abundance based upon the He/H ratio measured in regions of
hot, ionized gas (HII regions) found in metal-poor, dwarf emission-line
galaxies.  Using one sample and extrapolating to zero metallicity,
Olive and Steigman \cite{os} infer
$Y_P = 0.232 \pm 0.003{\rm\,(stat)} \pm 0.005{\rm \,(sys)}$;
using a new sample of objects Izotov et al \cite{izotovetal} infer
$Y_P = 0.243 \pm 0.003 {\rm \,(stat)}$.  Both data sets are
shown in Fig.~\ref{fig:he4}.  In brief, the
current situation is ambiguous, both as to the primeval $^4$He
abundance and as to the consistency of the big-bang prediction.

There has been much debate about $^4$He; there is a general consensus that
systematic error is the limiting factor at present.
Many effects have to be considered to achieve the desired accuracy:
corrections for doubly ionized $^4$He and neutral $^4$He have to be
made; absorption by dust and by stars have to be accounted for; collisional
excitation must be accounted for; potential
systematic errors exist in the input atomic physics;  and extrapolation
to zero metallicity must be made in the absence of a well motivated model.
(Regarding the last point, because there is more than one post-big-bang
source of $^4$He, the relationship between $Y_P$ and metallicity is
almost certainly not single-valued.)

Olive and Steigman argue that the systematic
error is no larger than 0.005, and their estimate of the primordial
$^4$He abundance is discrepant with the prediction based upon deuterium.
(Hata et al \cite{hataetal} have even gone so far as to argue
for a crisis; for another view see Ref.~\cite{cst3}.)
Others, including Pagel, Skillman, Sasselov and
Goldwirth, believe that the current systematic error budget is larger --
more like 0.010 or 0.015 -- in which case the discrepancy is at
most two sigma.  And of course, the Izotov et al value for $Y_P$
is consistent with the big-bang prediction.

Turning to the data themselves; the two samples are in general
agreement, except for the downturn at the lowest
metallicities which is seen in the data analyzed by Olive and Steigman.
(Skillman has recently also expressed concern about the use of the
lowest metallicity object, IZw18 \cite{skillman}.)  Visually, the data
make a strong case for a primordial $^4$He abundance that
is greater than 0.22 and less than about 0.25.

\begin{figure}[t]
\centerline{\psfig{figure=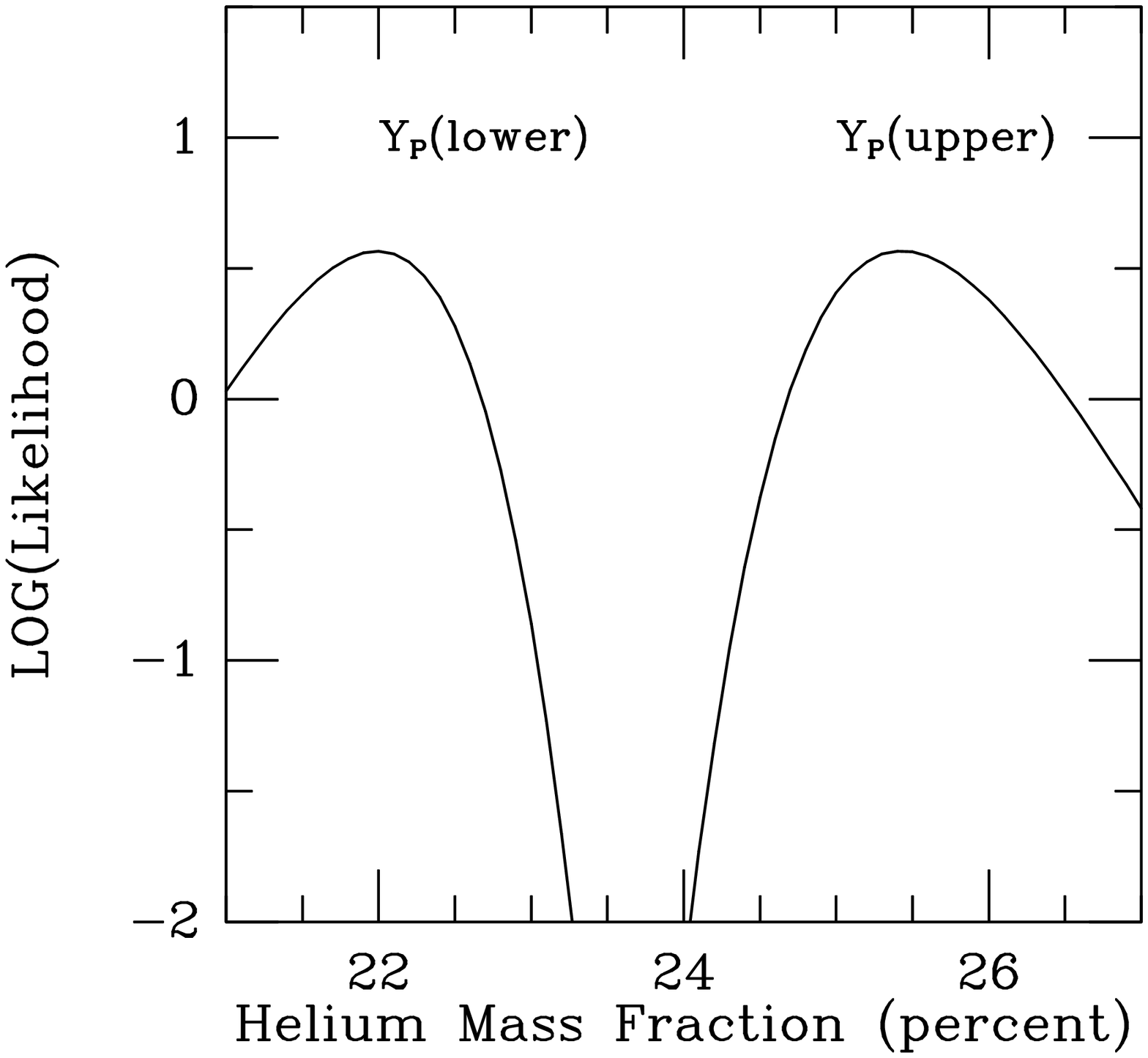,width=5in}}
\caption{Likelihood functions (unnormalized) for the
conservative lower limit to $Y_P$ (left) and conservative
upper limit to $Y_P$ (right).  These results are based
upon the HII regions in the sample analyzed by Olive and Steigman with
metallicity O/H$\le 10^{-4}$; qualitatively similar results
obtain for different metallicity cuts and for the Izotov et al sample.}
\label{fig:bayes}
\end{figure}

To be more quantitative about this statement and to derive
very conservative upper and lower bounds to $Y_P$, we have carried
out a nonparametric Bayesian analysis which makes minimal assumptions
about systematic error and the relationship between $Y_P$ and
metallicity.
We write the $^4$He abundance of a given object as $Y_i = Y_P + \Delta Y_i$.
To obtain a lower bound to $Y_P$ we take a flat prior for $\Delta Y_i$,
$0< \Delta Y_i < 1-Y_P$, which accounts
for the fact that stellar contamination increases the
$^4$He abundance.  To obtain an upper bound to $Y_P$ we take
a different flat prior, $-Y_P <\Delta Y_i <0$, which accounts for possible
systematic error that might lead to an underestimation of the $^4$He
abundance in an object.  The likelihood distributions for the lower bound
to $Y_P$ (first method) and for the upper bound (second
method) are shown in Fig.~\ref{fig:bayes}.
The 95\% confidence intervals for the two bounds are:
\begin{eqnarray}
{\rm Conservative\ Lower\ Limit:} & \qquad & Y_P({\rm lower}) =
        0.220^{+0.006}_{-0.012} \nonumber\\
{\rm Conservative\ Upper\ Limit:} & \qquad & Y_P({\rm upper}) =
        0.253^{+0.015}_{-0.005} \nonumber\\
\end{eqnarray}
While one certainly would like to do better in pinning down $Y_P$,
this very conservative analysis illustrates the strength of the
case for a primeval $^4$He mass fraction between 22\% and 25\%.
(Recently, Hogan et al have carried out a similar analysis to derive a
similar lower limit to $Y_P$ \cite{hoganetal}.)

At the moment $^4$He is a loose end.  Once the systematic
uncertainties are under control, $^4$He has an important
role to play in the high-precision era, as a test of the consistency of BBN.
Skillman and others are talking about a new assault on $Y_P$ --
putting together a larger, more homogeneous set of low-metallicity
galaxies in order to better understand, and hopefully reduce, systematic error.
It will be interesting to
see if it turns out to be the loose end that unravels the tapestry
or if it is woven back into the tapestry.
The resolution of the $^4$He problem could even involve new physics --
a short lived tau neutrino of mass greater than a few MeV could lower the
prediction for $Y_P$ by as much as $\Delta Y = 0.012$ -- but it
is certainly premature to give much weigh to this possibility.

\section{A New Test of The Standard Theory}

Almost overnight, the discovery of the
Cosmic Background Radiation transformed cosmology from the
realm of a handful of astronomers to a branch of physics.
Moreover, it was considerations of big-bang nucleosynthesis
that led Gamow, Peebles and others to predict the existence of
the CBR \cite{bbnstory}.  In the next decade, the CBR will likely
return the favor by providing an important new check of big-bang
nucleosynthesis.

The BBN test is part of a larger program to harvest the wealth of
information about the early Universe that is encoded in the
anisotropies of the CBR.  The anisotropy of the CBR is most naturally
described by its multipole decomposition
\begin{equation}
{\delta T(\theta ,\phi )\over T} = \sum_{lm} a_{lm}Y_{lm}(\theta ,\phi )\,.
\end{equation}
For a theory like inflation, where the underlying density perturbations that
lead to the anisotropy are gaussian, all information is encoded
in the variance of the multipole amplitudes.  (The multipoles are
gaussian distributed with zero mean, with the {\it rms} temperature difference
between directions on the sky separated by angle $\theta$ given roughly
by $\sqrt{l(l+1)C_l /2\pi }$ with $l \approx 180^\circ /l$.)
The angular power spectrum, $C_l \equiv \langle
|a_{lm}|^2\rangle$, depends not only on the spectrum of
density perturbations, but also upon cosmological parameters,
including the baryon density.  

\begin{figure}[t]
\centerline{\psfig{figure=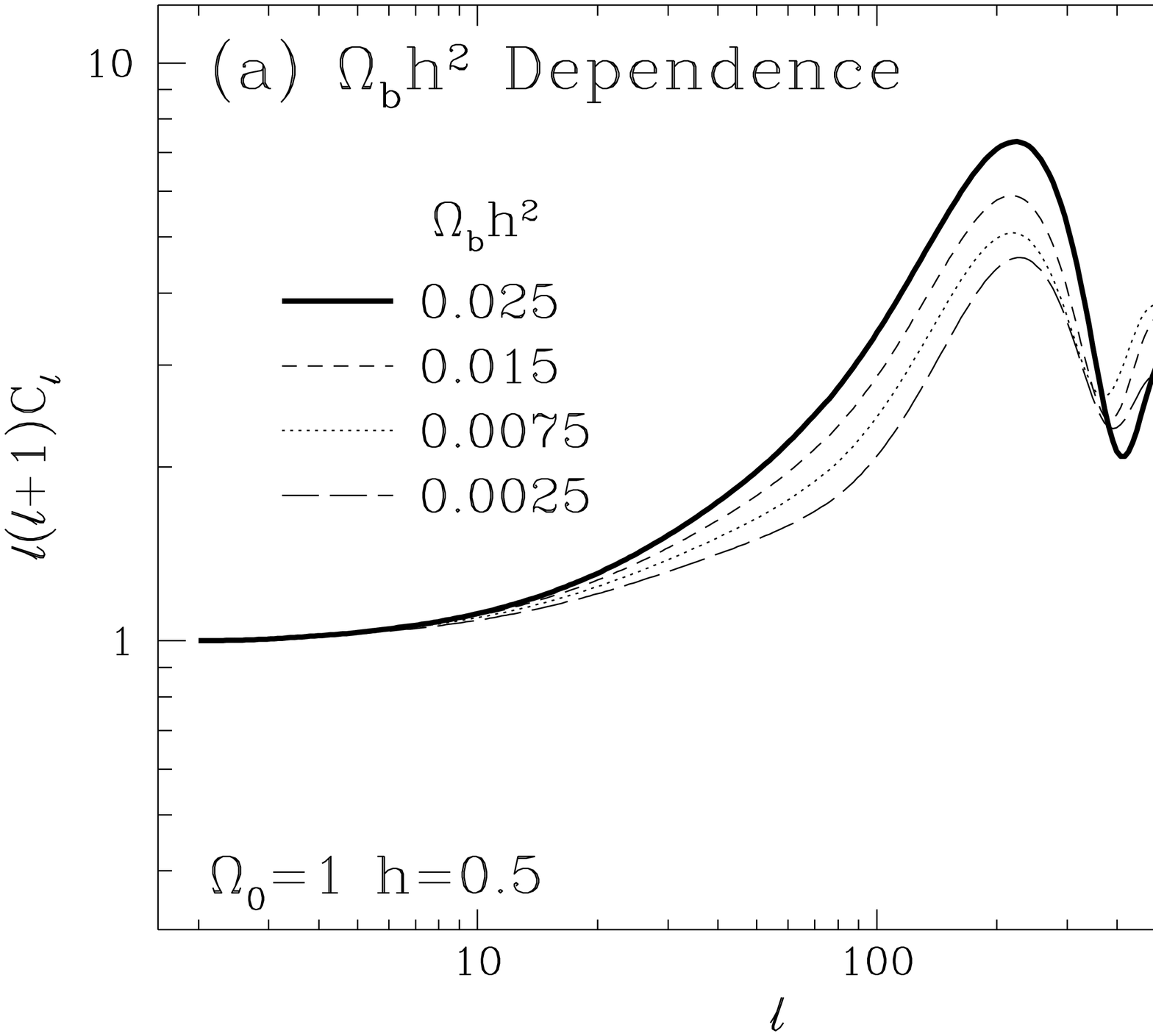,width=4in}}
\caption{Dependence of the angular power spectrum of CBR anisotropy
on baryon density for cold dark matter models (courtesy of Martin White).}
\label{fig:baryoncbr}
\end{figure}

The angular power spectrum, shown in Fig.~\ref{fig:baryoncbr},
is characterized by a featureless (Sachs-Wolfe) plateau from
$l=2$ to $l\sim 100$; and a series of (acoustic or Doppler) peaks
and valleys from $l=200$ to $l\sim 2000$; for $l\gg 2000$ anisotropy
is strongly damped by photon diffusion which smears out anisotropy
on smaller scales \cite{wayne}.
The plateau arises due to differences in the gravitational potential
on the last scattering surface (Sachs-Wolfe effect).  The peaks
and valleys arise due to photon-baryon acoustic oscillations driven
by gravity, and their amplitudes and spacings depend upon the contribution
of baryons to the matter density (see Fig.~\ref{fig:baryoncbr}).

When the two new satellite-experiments, NASA's MAP to be launched
in August 2000 and ESA's Planck to be launched in 2005, map
the sky with angular resolution of $0.1^\circ$ (or better), they
will determine the variance of about 2500 multipoles to an
accuracy essentially limited by sky coverage and sampling variance.
From this it should be possible to determine precisely a
number of cosmological parameters, including the total energy
density ($\Omega_0$) and the fraction of critical density contributed by
matter ($\Omega_M$), a cosmological constant ($\Omega_\Lambda$), and
neutrinos ($\Omega_\nu$); the Hubble constant ($H_0$); the power-law
index of the spectrum of density perturbations ($n$) and deviation
from an exact power law ($dn/d\ln k$); the contribution of gravitational
waves to the anisotropy (tensor to scalar ratio $T/S$); and the baryon density
($\Omega_Bh^2$).  In particular, the baryon density should ultimately
be determined to a precision of around 5\% \cite{learnfromcbr}.

Even before MAP flies a host of balloon-borne and
ground-based experiments (e.g., CBI, MAXIMA, VCA, VSA, BOOMERANG, QMAT,
and TOPHAT) will cover a significant fraction of the sky with
angular resolution of less than one degree.  These experiments
may be able to delineate the first two or three acoustic peaks and
thereby determine the baryon density to 25\% or so.

Certainly within a decade, and probably much sooner, there will be
an independent, high-precision determination of the baryon density
which is based on very different physics -- gravity-driven, acoustic
oscillations of the photon-baryon fluid when the Universe was around
300,000 years of age.  If this determination of the baryon density
agrees with that based upon big-bang nucleosynthesis it will be an
impressive confirmation of the standard cosmology as well as
the standard cosmology and general relativity.

\section{Probing Particle Physics with New Precision}

\begin{figure}[t]
\centerline{\psfig{figure=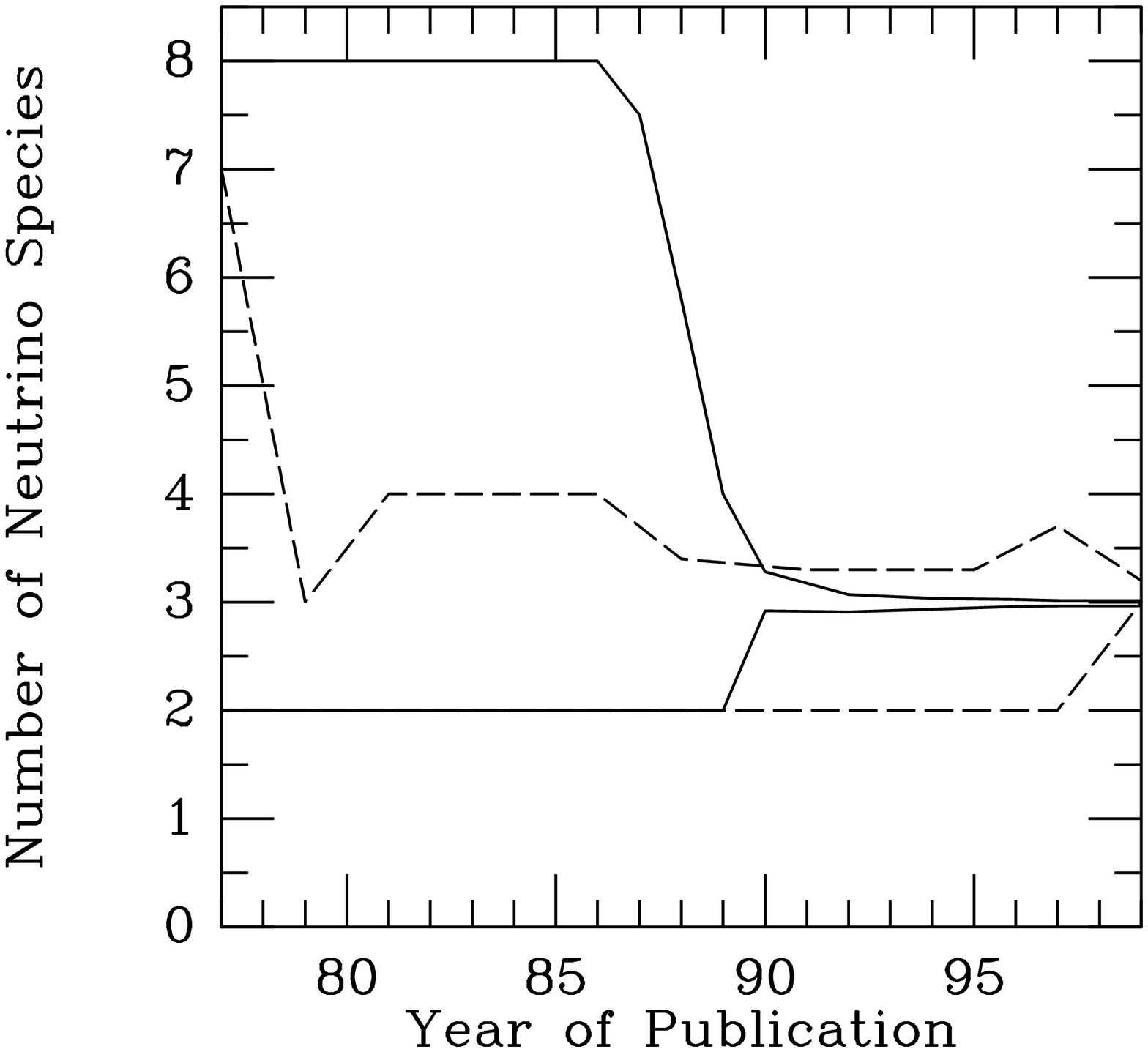,width=5in}}
\caption{Cosmological (broken curve) and laboratory (solid curve)
limits (95\% cl) to the number of neutrino species.  An
ultimate cosmological limit of 3.1 neutrino species has been ``anticipated.''}
\label{fig:nucount}
\end{figure}

For almost two decades, big-bang nucleosynthesis has also been a powerful
probe of fundamental physics, best illustrated by
the BBN limit to the number of light neutrino species.  In 1977
Gunn, Schramm and Steigman argued that big-bang helium production
set a limit of less than seven light neutrino species \cite{gss};
by 1980 the limit had been refined to less than four neutrino
species.  Not until the $Z^0$-factories
at SLAC and CERN came on line in 1989 did the laboratory limit become
competitive (see Fig.~\ref{fig:nucount}).  Today, the LEP limit
based upon the shape of the $Z^0$ resonance stands at $N_\nu =
2.989\pm 0.024$ (95\% cl), a truly impressive achievement.

While it is unlikely that the big-bang nucleosynthesis limit
will ever achieve such precision, it will improve significantly
when the baryon density is determined accurately.
Moreover, the cosmological and laboratory limits are complementary:
The neutrino limit based upon the shape of the $Z^0$ counts the
number of particle species that are less massive than half the $Z^0$ mass,
weighted by their coupling to the $Z^0$.  BBN constrains
the energy density contributed by relativistic particle species
around the time of primeval nucleosynthesis and thus is sensitive to
any particle species lighter than about $1\MeV$.
Historically, both have been expressed as a limit to the
number of neutrino species.

Let's quickly review the physics of the big-bang nucleosynthesis
limit.  The amount of $^4$He synthesized depends strongly
upon the expansion rate,
which determines the neutron fraction at the epoch of nucleosynthesis,
and weakly upon the baryon density, which determines reaction
rates (see Fig.~\ref{fig:bbnbig}).   A faster expansion rate
leads to more $^4$He production because the neutron fraction
freezes out at a higher value; higher baryon density leads to
more production of $^4$He as nuclear reactions begin earlier
when the neutron fraction is higher.  The expansion rate itself
is determined by the energy density in relativistic particles,
parameterized by the effective number of massless particle species,
\begin{equation}
g_* = \sum_{m\la 1\MeV}^{\rm Fermi}g_i(T_i/T)^4
         +{7\over 8}\sum_{m\la 1\MeV}^{\rm Bose}g_i(T_i/T)^4\,,
\end{equation}
where $T_i$ is the temperature of species $i$.  A species that
interacts more weakly than neutrinos can have a lower temperature
than the plasma temperature \cite{kt}.
The particles in the standard model contribute 10.75 to the sum, with each
neutrino species contributing 1.75.

In the absence of precise knowledge of the baryon density and
the measured primeval $^4$He abundance, setting a big-bang limit
requires a {\it lower limit} to the baryon density and an upper
limit to the value of the primeval $^4$He abundance.
For more than a decade the lower limit to the baryon density
was based upon the upper limit to the big-bang production of D+$^3$He.
The upper limit to the primeval production of $^4$He was assumed to
be 25\% (and sometimes as low as 24\%).  Much progress has
been made on the former  -- the provisional value of the
primeval deuterium abundance pegs the baryon density to a precision of
20\% at a value that is a factor of three above the previous lower limit.

Pinning down the baryon density improves the big-bang neutrino
limit significantly.  A recent Bayesian analysis assuming a primeval
$^4$He abundance $Y_P=0.242\pm 0.003$ gave the following 95\%
credible intervals for $N_\nu$:  $N_\nu = 3.0 - 3.7$, assuming
the D+$^3$He lower bound to the baryon density (constrained by
metal production), and $N_\nu =
3.0 - 3.2$, assuming (D/H)$_P = (2.5\pm 0.75 )\times 10^{-5}$
\cite{cst4} (in both cases the prior $N_\nu \ge 3$ was enforced).

The determination of the baryon density from the primeval deuterium
abundance will have a similarly dramatic impact on the big-bang limit
as the commissioning of the $Z^0$ factories did on the laboratory limit.
When the baryon density is known to a precision of 5\%
and when the systematic uncertainties in the $^4$He abundance are reduced,
an upper limit as precise as 3.1 neutrino species is likely.
Together, the cosmological and laboratory neutrino limits work hand in
hand to constrain new physics.

\section{Concluding Remarks}

Big-bang nucleosynthesis is a cornerstone
of the standard cosmology.  Together with the CBR it provides
compelling evidence that the early Universe
was hot and dense.  This opened the door to the study of the
earliest moments and helped to forge the symbiotic relationship
between particle physics and cosmology.  The inner space -- outer
space connection has led to very interesting and attractive ideas
about the earliest moments, including inflation and cold dark matter.
These ideas are now being tested by a host of experiments and observations
and in process a new window to fundamental physics is being opened
\cite{physworld}.

For more than two decades BBN has also provided the best
determination of the baryon mass density, which has led
to three important conclusions:  baryons cannot provide the
closure density; most of the baryons are dark and most of the
dark matter is nonbaryonic.

As we have tried to emphasize and illustrate, the pegging of
the baryon density by a determination of the primeval deuterium abundance
will advance BBN to a new, precision era.  The harvest to come
is impressive:  An accurate determination
of a fundamental parameter of cosmology; light-element tracers
to study Galactic and stellar chemical evolution; and new precision in
probing fundamental physics.  Finally, there are two important
tests of BBN on the horizon:  a precision check of the predicted
primeval $^4$He abundance by new measurements;
and a comparison of the BBN value for the baryon density with
that derived from CBR anisotropy.

\paragraph{Acknowledgments.}
This work was supported by
the DoE (at Chicago and Fermilab), and by NASA (at Chicago and
Fermilab by grant NAG 5-2788), and by the NSF (at Chicago).

\end{document}